\documentclass{INTERSPEECH2023}
\usepackage{multirow}
\usepackage{cite}
\usepackage{tabularx}
\usepackage{lipsum}

\usepackage{gensymb}
\usepackage[table,xcdraw]{xcolor}

\title{FN-SSL: Full-Band and Narrow-Band Fusion for Sound Source Localization}
\name{Yabo Wang$^{1,2,\dag}$, Bing Yang$^{2,\dag}$, Xiaofei Li$^{2,*}$}
\address{
  $^1$Zhejiang University, Hangzhou, China\\
  $^2$Westlake Institute for Advanced Study \& Westlake University , Hangzhou, China}
\email{wangyabo@westlake.edu.cn, yangbing@westlake.edu.cn, lixiaofei@westlake.edu.cn}

\begin{document}

\maketitle

\newcommand\blfootnote[1]{%
\begingroup
\renewcommand\thefootnote{}\footnote{#1}%
\addtocounter{footnote}{-1}%
\endgroup
}
\begin{abstract}
Extracting direct-path spatial features is critical for sound source localization in adverse acoustic environments. This paper proposes a full-band and narrow-band fusion network for estimating direct-path inter-channel phase difference (DP-IPD) from microphone signals. The alternating full-band and narrow-band layers are responsible for learning the full-band correlation and narrow-band extraction of DP-IPD, respectively. Experiments show that the proposed network noticeably outperforms other advanced methods on both simulated and real-world data. 
\end{abstract}
\noindent\textbf{Index Terms}: Moving sound source localization, full-band, narrow-band,  direct-path, inter-channel phase difference
\vspace{-0.4em}
\section{Introduction}
Sound source localization (SSL) estimates the position of sound sources using the multi-channel sound recordings. It is widely used in applications such as video conferencing and robot audition, and also employed to boost the speech enhancement and separation performance \cite{Lee2016DNNBasedFE, Chazan2019MultiMicrophoneSS}. 
Conventional localization methods \cite{Zhang2010ATM, Knapp1976TheGC, 4967888} usually first estimate the spatial features, such as time delay, inter-channel phase/level difference (IPD/ILD) and relative transfer function (RTF), and then build the feature-to-location mapping. 
Mapping spatial features to source location is straightforward if the spatial features are clean. The localization features correlate to the direct-path signal propagation, however, the direct-path signal is often contaminated by noise and reverberation in real-world environments, which significantly degrade the accuracy of feature estimation. 
\blfootnote{\dag  Equal Contribution, * Corresponding Author. This work was supported by Zhejiang Provincial Natural Science Foundation of China under Grant 2022XHSJJ008, and Postdoctoral Science Foundation of China under Grant 2022M722848.}

Recently, deep learning based localization methods have been widely studied 
\cite{grumiaux2022survey}. 
They treat the localization task as feature/location regression \cite{DiazGuerra2020RobustSS,9413923} or location classification problem \cite{Chakrabarty2018MultiSpeakerDE, 8086216}. 
The commonly adopted network architectures for moving SSL include Convolutional Neural Networks (CNN) \cite{DiazGuerra2020RobustSS, DiazGuerra2022DirectionOA} and Convolutional Recurrent Neural Networks (CRNN) \cite{Yang2022SRPDNNLD, Nguyen2021SALSALiteAF, yang2021learning, Han_KU_task3_report, Adavanne2018_JSTSP}. 
Convolutional layers are used to extract local spatial information, and RNNs capture the long-term temporal context of spatial information. 
The input of deep learning models can be in signal-level such as time-domain signal and  magnitude/phase spectrograms, or in feature-level such as IPD \cite{Nguyen2021SALSALiteAF}, generalized cross-correlation (GCC) function \cite{Knapp1976TheGC} and spatial spectrum \cite{DiBiase2001RobustLI, DiazGuerra2020RobustSS}. 
 Based on previous related works, it is known that the direct-path features can be estimated in both narrow-band and full-band. \textbf{1) Narrow-band processing}: the signals (with reverberation) are often formulated in the short-time Fourier transform (STFT) domain, and the STFT frequencies can be processed independently, as the random signals and reverberation are statistically independent between frequencies. In \cite{7533416,li2019online2}, the direct-path feature is estimated based on sub-band channel identification. According to the precedence effect \cite{litovsky1999precedence} that the direct-path signal arrives first relative to the reflections, the direct-path dominated frames are detected by performing coherence test in \cite{Mohan2008LocalizationOM} and by performing direct-path dominance test in \cite{Nadiri2014LocalizationOM}. 
\textbf{2) Full-band processing}: 
direct-path features of the directional sound source are highly correlated across frequencies, such as the IPD of all frequencies correspond to (and are linearly related to) the time delay of arrival (TDOA) \cite{gannot2017consolidated}. This property is largely exploited by the methods \cite{Nguyen2021SALSALiteAF} that performs deep regression from the noisy IPD to position or clean IPD.

To fully leverage the narrow-band and full-band information mentioned above, this work proposes a full-band and narrow-band fusion network for direct-path IPD (DP-IPD) estimation. 
The network takes as input the multichannel microphone signals, and predicts the frame-wise DP-IPD for online SSL. This follows the principle of the SSL methods presented in \cite{Yang2022SRPDNNLD, yang2021learning}. Compared to processing the noisy IPD  \cite{Nguyen2021SALSALiteAF} or the noisy spatial spectrum \cite{DiazGuerra2020RobustSS},   
directly processing the microphone signals is more effective to leverage the natural properties of noise and reverberation for suppressing them, such as the spatial-diffuseness of noise and late reverberation.
Specifically, the proposed full-band and narrow-band fusion network uses dedicated LSTM networks for full-band and narrow-band processing, respectively. Full-band (narrow-band) layers process the time frames (frequencies) independently, and all the time frames (frequencies) share the same network parameters. The input for full-band (narrow-band) layers is a sequence along the frequency axis (time axis) of one single time frame (frequency). The cascaded and alternated full-band and narrow-band layers respectively focus on learning the full-band correlation and narrow-band extraction of DP-IPD. We also design the respective skip connections along the full-band and narrow-band layers to avoid information loss. Experiments on both simulated and real-world data show that the proposed method noticeably outperforms the recently proposed SSL methods, and performs well under conditions with strong reverberation and noise.

The proposed full-band and narrow-band fusion network is motivated by the recently proposed speech enhancement networks \cite{tesch2022insights, Yang2022McNetFM}, in which one layer of full-band LSTM and one layer of narrow-band LSTM is cascaded for predicting the clean speech signal. This full-band and narrow-band fusion scheme shows a large performance superiority for speech enhancement, and is now becoming a new research trend. Speech enhancement and SSL are two entangled tasks, as they respectively estimate the direct-path speech signal and direct-path signal propagation. In this work, we redesign the full-band and narrow-band fusion network for SSL, named as FN-SSL network, mainly by extending the network to alternate between full-band and narrow-band layers and designing proper SSL target.   

\vspace{-0.4em}
\section{Method}
In an enclosed environment, a moving sound source at a direction $\theta$ is observed by two microphones. The microphone signals can be defined in the STFT domain as:
\begin{equation}
X_m(t, k)=A_m(t,k, \theta) S(t, k)+N_m(t, k),
\end{equation}
where $m \in[1, 2]$, $t \in[1, T]$ and $k \in[1, K]$ denote the indices of microphone, time frame and frequency, respectively. $X_m(t, k)$, $S(t, k)$ and $N_m(t, k)$ are the STFT
coefficients of microphone signal, source signal and noise, respectively. 
$A_m(t,k, \theta)$ is the room transfer function, as the Fourier transform of room impulse response (RIR). Note that this work considers moving sound source, thus the room transfer function is time-varying/dependent. 
\vspace{-0.4em}
\subsection{Learning targets}  
\subsubsection{Direct-path IPD}
The room transfer function can be decomposed as:
\begin{equation}
A_m(t,k, \theta)=A_m^{\mathrm{d}}(t,k, \theta)+A_m^{\mathrm{r}}(t,k, \theta),
\end{equation}
where $A_m^{\mathrm{d}}(t,k, \theta)$ and $A_m^{\mathrm{r}}(t,k, \theta)$ represent the direct-path and reverberation parts, respectively. 
The source direction is only relevant to the direct-path propagation. In \cite{7533416,li2019online2}, the direct-path relative transfer function (DP-RTF), defined as $ B^{\mathrm{d}}(t,k, \theta)={A_2^{\mathrm{d}}(t,k, \theta)} / {A_1^{\mathrm{d}}(t,k, \theta)}$, is taken as a localization feature, which encodes the direct-path IPD and ILD within its phase and amplitude, respectively. As IPD is more discriminative than ILD, we only take the direct-path IPD, i.e. $\angle B^{\mathrm{d}}(t,k, \theta)$, as the localization feature of this work. DP-IPD corresponds to the TDOA of the direct-path propagation from source to two microphones. The direct-path transfer functions of a given direction $\theta$ can be obtained by applying Fourier transform to the direct-path impulse responses (or head-related impulse responses for binaural localization). 

FN-SSL takes as input the dual-channel noisy signal, i.e. $X_m(t, f)$, and outputs/predicts the DP-IPD features. To enable the optimization of real-value networks, the learning target (for frame $t$) is set as the real and imaginary parts of complex-valued DP-IPD (concatenated along frequencies) \cite{yang2021learning,Yang2022SRPDNNLD} as $[\cos \angle B^{\mathrm{d}}(t,1, \theta), \sin \angle B^{\mathrm{d}}(t,1, \theta),..., \cos \angle B^{\mathrm{d}}(t,K, \theta),  \sin  \angle \\ B^{\mathrm{d}}(t,K, \theta)]^T,$ 
where $^T$ denotes vector transpose. The output activation layer is set as \emph{tanh} to predict DP-IPD. The mean squared error is used for training.
At inference, the inner product between the predicted DP-IPD vector of one frame and the DP-IPD vector of candidate directions are computed, and the candidate direction with the largest inner product is taken as the localization result. The candidate directions are sampled in the whole localization space.

\begin{figure}[t]
    \centering
    \includegraphics[width=0.77\linewidth]{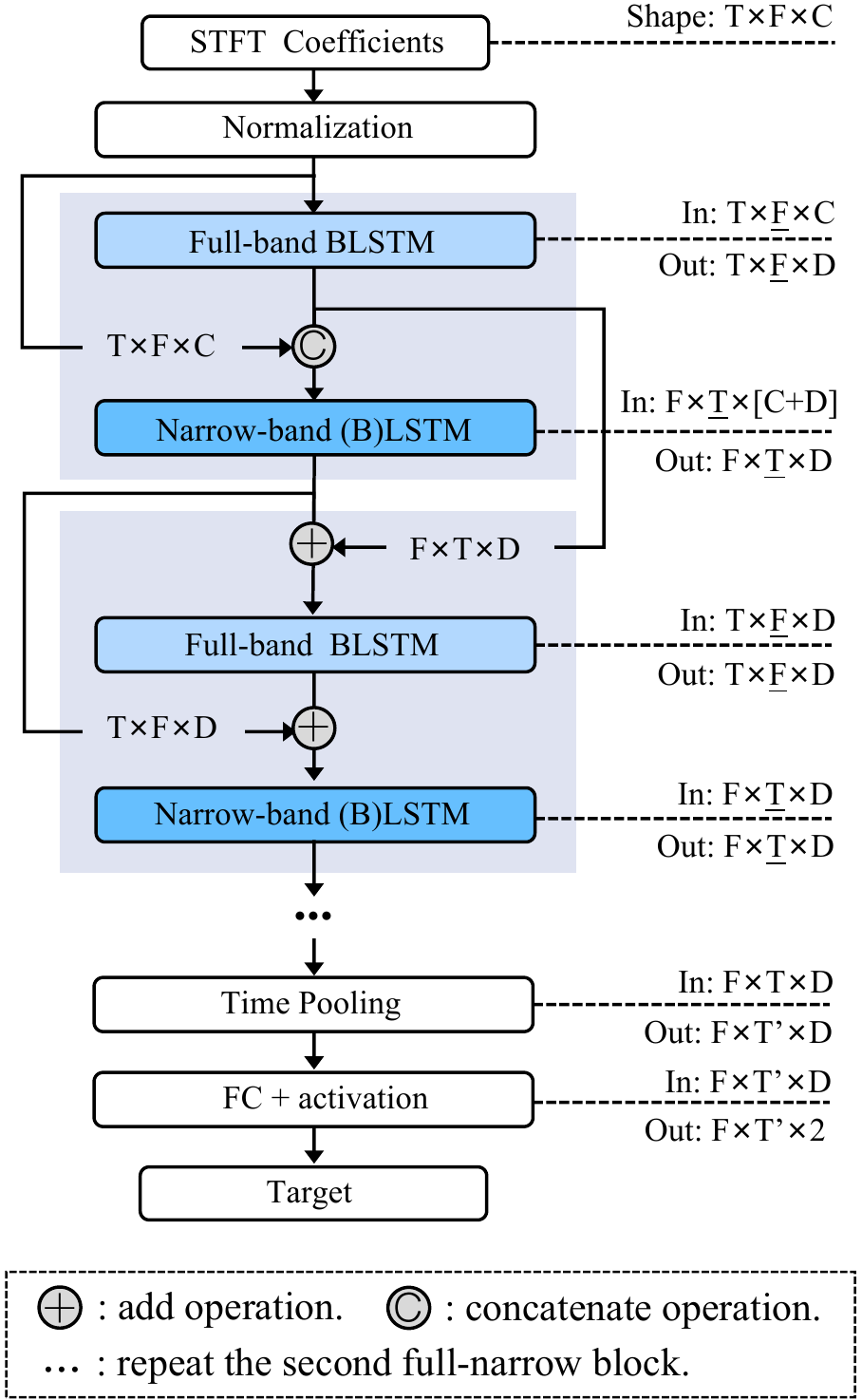}
    \vspace{-0.5em}
    \caption{Model architecture of the proposed FN-SSL network. A full-band layer plus a narrow-band layer form a full-narrow block. Each LSTM layer processes a number of independent feature sequences. The data organization is in the format of: number of sequences $\times$ \underline{sequence length} $\times$ feature dimension.}
    \label{fig:method}
   \vspace{-2em}
\end{figure}
\vspace{-0.4em}
\subsubsection{Optional targets}\label{sec:otarget}
Although the proposed network is designed for DP-IPD estimation, it can also be directly used in the framework of location classification or regression to perform end-to-end SSL, by simply replacing the DP-IPD target with location class or location coordinate. For location classification, the localization space is divided into a number of location classes, the output  activation layer is set to Softmax, and the cross entropy loss is used for training. For location regression, the target is set as the $\ell_2$-normalized coordinate, the network output is also $\ell_2$-normalized, and the mean squared error is used for training.
\vspace{-0.2em}
\subsection{Network Architecture}
The structure of the proposed FN-SSL network is shown in Fig. \ref{fig:method}. The real and imaginary parts of the STFT coefficients are taken as the network input, and thus the number of input channels $C$ is 2 times the number of microphones. The input is processed by alternating full-band and narrow-band LSTM layers. One full-band layer plus one narrow-band layer is called a full-narrow block. Fig. \ref{fig:method} shows a network with two full-narrow blocks, and more blocks can be easily added by repeating the second block. The output of the full-narrow blocks is passed to a average pooling module to compress the frame rate, and then fed to a fully connected (FC) layer (with \emph{tanh} activation) to convert to the desired output dimension. 
                                              

\vspace{-0.4em}
\subsubsection{Full-band BLSTM layer}


The full-band BLSTM layers process the time frames independently, and all the time frames share the same network parameters. The input is a sequence along the frequency axis of one single time frame: 
\begin{equation}
    H^f(t) = (\mathbf{h}^\textit{f}(t,1), \dots,\mathbf{h}^\textit{f}(t,k), \dots , \mathbf{h}^\textit{f}(t,K)),
\end{equation}
where the superscript $^f$ indicates full-band layer, $\mathbf{h}^\textit{f}(t,k)\in \mathcal{R}^{D\times1}$ represents the hidden vector of a time-frequency (TF) bin.  $\mathbf{h}^\textit{f}(t,k)$ is the microphone signals for the first full-band layer, while it is the output of the previous narrow-band layer for other full-band layers. The full-band layers focus on learning the inter-frequency dependencies of spatial/localization cues. DP-IPD of different frequencies has a strong correlation, as they are all derived from the same TDOA. In addition, the spatial/localization cues of those frequencies with low direct-path energy can be well predicted with the help of other frequencies. The full-band layers do not learn any temporal information, which is left for the following narrow-band layers. 

\vspace{-0.3em}
\subsubsection{Narrow-band (B)LSTM layer}
The narrow-band (B)LSTM layers process the frequencies independently, and all the frequencies share the same network parameters. The input is a sequence along the time axis of one single frequency: 
\begin{equation}
    H^n(k) = (\mathbf{h}^\textit{n}(1,k), \dots,\mathbf{h}^\textit{n}(t,k), \dots , \mathbf{h}^\textit{n}(T,k)),
\end{equation}
where the superscript $^n$ indicates narrow-band layer. The input hidden vector $\mathbf{h}^\textit{n}(t,k)\in \mathcal{R}^{D\times1}$ is the output vector of the previous full-band layer. Estimating the direct-path localization features in narrow-band has been studied in many conventional methods, such as by channel identification in \cite{7533416}, coherence test in \cite{Mohan2008LocalizationOM}, direct-path dominance test in \cite{Nadiri2014LocalizationOM}. The proposed narrow-band layers focus on exploiting these narrow-band inter-channel information. In addition, DP-IPD is time-varying for moving sound source, and the narrow-band layers learn the temporal evolution of DP-IPD as well.

\vspace{-0.3em}
\subsubsection{Skip connections}  
The full-band and narrow-band layers are designed to focus on their respective information. The narrow-band information may be lost after a full-band layer, and vice versa. Therefore, we add  skip connections to avoid the information loss. As shown in Fig.~\ref{fig:method}, the narrow-band microphone signals are concatenated to the output of the first full-band layer as the input of the first narrow-band layer. Afterwards, we add the output of the previous full-band layer (narrow-band layer) to the input of the next full-band layer (narrow-band layer). 
\vspace{-0.3em}
\subsubsection{Model causality}
The proposed network can be easily implemented for both offline or online SSL, by setting the narrow-band LSTMs to be bidirectional or unidirectional, respectively. 
To make the model easier to optimize, Laplace normalization is performed on the network input as $X_m(t, f) / \mu(t)$, where $\mu(t)$ is a normalization factor. For offline SSL, $\mu(t)$ is computed as $\frac{1}{TK} \sum_{t=1}^T \sum_{k=1}^{K}\left|X_m(t, f)\right|$. For online SSL, $\mu(t)$ is recursively calculated as $\mu(t)=\alpha \mu(t-1)+(1-\alpha) \frac{1}{K} \sum_{k=1}^{K}\left|X_m(t, k)\right|$ \cite{Yang2022McNetFM} to ensure the causality of the method. Here, $\alpha={(L-1)}/{(L+1)}$ denotes the smoothing weight of the historical time frames, and $L$ represents the length of smoothing window.

\vspace{-0.3em}
\section{Experiments}
\subsection{Dataset and configurations}

The proposed method is evaluated on both simulated and real-world datasets. We consider to use two microphones to localize the direction of arrival (DOA) for 180$^\circ$ azimuth angles.

\textbf{Simulated dataset.} Microphone signals are obtained by convoluting RIRs with speech source signals. The speech signals randomly selected from the training, dev and test sets of the LirbriSpeech corpus \cite{Panayotov2015LibrispeechAA} are used for our model training, validation and test, respectively. RIRs are generated using the gpuRIR toolbox \cite{DiazGuerra2018gpuRIRAP}. The room reverberation time (RT60) is randomly set in the range of [0.2, 1.3] s. The room size is randomly set in the range from 6×6×2.5 m to 10×8×6 m. The moving trajectory of speech sources are generated according to \cite{DiazGuerra2020RobustSS}, and each has a fix height. Two microphones with a distance of 8 cm are randomly placed in the room, which are in the same horizontal plane as the sound source. 
We employ white, babble, and factory noises from the NOISEX-92 database \cite{Varga1993AssessmentFA} as noise sources, and create the diffuse sound field following \cite{Habets2008GeneratingNM}. Generated diffuse noise signals are added to the clean sensor signals according to a signal-to-noise ratio (SNR) randomly selected from -5 dB to 15 dB. The number of utterances for training, validation and test are 166,816, 992, and 5,000, respectively. 

\textbf{Real-world dataset.} We evaluate our method on the task 3 and 5 of the LOCATA dataset \cite{Lllmann2018TheLC}. The room size is 7.1×9.8×3 m and reverberation time is 0.55 s. The microphones 6 and 9 of the DICIT array \cite{Brutti2010WOZAD} are used for evaluation, which have the same configuration as the simulated microphone array. Note that there are four recordings whose azimuth angles are out of the range of [0, 180$^\circ$] in task 5, and we don’t consider them in our experiments. The models trained with the simulated dataset are directly used for testing on the LOCATA dataset.

\textbf{Configurations.}
The sampling rate is 16 kHz. The window length of STFT is 512 samples (32 ms) with a frame shift of 256 samples (16 ms), and Hanning window is used. The length of audio used for training is 4.79 s. We concatenate the real and imaginary parts of the STFT coefficients as the network input, and the channel of input is $C = 2M$, where $M = 2$ is the number of microphones. The time frames during training are pooled from 298 to 24. The output dimension of every (B)LSTM layers are all set to 256. 
MSE is used as the loss function.
During training, Adam is used as the optimizer, and the batch size is set to 16. The initial learning rate is set to 0.001, and the learning rate is exponentially decayed with a decaying factor of 0.8988. We train the model for 15 epochs. 

Mean Absolute Error (MAE) and Localization Accuracy (ACC) are used as the evaluation metrics. ACC ($N^{\circ}$) means the ratio of frames with localization error less than $N^{\circ}$.  The resolution of candidate azimuth angles is set to 5$\degree$ for the simulated dataset and 2.5$\degree$ for the real-world dataset.  Only the non-silent frames are used for performance evaluation. Code is available on the website\footnote{https://github.com/Audio-WestlakeU/FN-SSL}. 

\textbf{Comparison methods.}
We compared with the following moving sound source localization methods: \textbf{(1) Cross3D}  \cite{DiazGuerra2020RobustSS} takes the SRP-PHAT spatial spectrum as input, and uses a causal 3D CNN network to perform moving sound source tracking.
\textbf{(2) IcoCNN} \cite{DiazGuerra2022DirectionOA} uses an Icosahedral CNN to extract localization feature from the SRP-PHAT spatial spectrum, and uses a casual 1D CNN to combine the temporal context for moving sound source tracking. 
\textbf{(3) SRP-DNN} \cite{Yang2022SRPDNNLD} is a causal CRNN network for estimating the DP-RTF feature of moving sound sources. \textbf{(4) SELDnet} is the CRNN baseline of DCASE22 \cite{Adavanne2018_JSTSP}. The input feature is the frequency normalized IPD concatenated with the magnitude spectrum. \textbf{(5) SE-Resnet} \cite{Han_KU_task3_report} is a top-ranked method of DCASE22 which uses a squeeze-and-excitation residual network (as encoder) and a Gated Recurrent Unit network (as decoder). \textbf{(6) SALSA-Lite} \cite{Nguyen2021SALSALiteAF} uses the frequency normalized IPD concatenated with the magnitude spectrum as the network input, as is done in SELDnet, and uses a ResNet-GRU network. SELDnet, SALSA-Lite and SE-Resnet are proposed for joint sound event detection and localization, and we only use their localization branch. All of these comparison methods are reproduced and trained using the same dataset as the proposed method. We modify the time pooling kernel size of them to ensure the same output frame rate for all methods.

\begin{table}[t]
\footnotesize
\caption{Ablation studies of FN-SSL.}
\vspace{-1em}
\renewcommand\arraystretch{1.1}
\tabcolsep0.06in
\begin{tabular}{lcccc}
\hline
\multicolumn{1}{c}{Methods}                         & \# Params.{[}M{]} & ACC (5$\degree$) {[}\%{]} & MAE {[}°{]}   \\ \hline
\multicolumn{1}{l}{Block$\times$1}                         & 0.6               & 87.0             & 3.0          \\
\multicolumn{1}{l}{Block$\times$2}                         & 1.6               & 90.7             & 2.5          \\
\quad with location class.                & 1.7                  & 90.1                 & 2.5             \\
               \quad with location regress.               & 1.6                    & 87.6                  & 2.7                \\               
\rowcolor[HTML]{EFEFEF} 
\multicolumn{1}{l}{\cellcolor[HTML]{EFEFEF}Block$\times$3} & 2.5               & \textbf{91.4}    & \textbf{2.4} \\
               \quad w/o skip.                       & 2.5               & 90.5             & 2.6          \\ \hline
\end{tabular}
\label{ab_study}
\vspace{-1em}
\end{table}

\begin{table}[]
\caption{Results on simulated data.}
\vspace{-1em}
\renewcommand\arraystretch{1.13}
\tabcolsep0.0055in
\scriptsize
\begin{tabular}{ccccccccccccc}
\hline
\multicolumn{2}{c}{}                                               &  &                                                                                 &  & \multicolumn{2}{c}{\begin{tabular}[c]{@{}c@{}}SNR = 0 dB \\ RT60 = 1 s\end{tabular}} &  & \multicolumn{2}{c}{\begin{tabular}[c]{@{}c@{}}SNR = 15 dB\\ RT60 = 0.2 s\end{tabular}} &  & \multicolumn{2}{c}{AVG.}                                                                                  \\ \cline{6-7} \cline{9-10} \cline{12-13} 
\multicolumn{2}{c}{}                                               &  &                                                                                 &  & ACC(5\degree)                                   & MAE                                      &  & ACC(5\degree)                                    & MAE                                       &  & ACC(5\degree)                                             & MAE                                                \\
\multicolumn{2}{c}{\multirow{-3.5}{*}{Methods}}                      &  & \multirow{-3}{*}{\begin{tabular}[c]{@{}c@{}}\# Params. \\ {[}M{]}\end{tabular}} &  & {[}\%{]}                                  & {[}°{]}                                  &  & {[}\%{]}                                   & {[}°{]}                                   &  & {[}\%{]}                                            & {[}°{]}                                            \\ \cline{1-2} \cline{4-4} \cline{6-7} \cline{9-10} \cline{12-13} 
                          & Cross3D \cite{DiazGuerra2020RobustSS}                               &  & 5.6                                                                             &  & 37.8                                      & 12.1                                     &  & 77.3                                       & 3.4                                       &  & 50.8                                                & 8.4                                                \\
                          & IcoCNN \cite{DiazGuerra2022DirectionOA}                                &  & 0.3                                                                             &  & 39.6                                      & 11.4                                     &  & 77.4                                       & 3.4                                       &  & 51.4                                                & 8.1                                                \\
                          & SRP-DNN \cite{Yang2022SRPDNNLD}                               &  & 0.8                                                                             &  & 80.4                                      & 3.8                                      &  & 98.3                                       & 1.7                                       &  & 86.9                                                & 3.0                                                \\
\multirow{-4}{*}{Online}  & \cellcolor[HTML]{EFEFEF}FN-SSL (prop.) &  & \cellcolor[HTML]{EFEFEF}2.5                                                     &  & \cellcolor[HTML]{EFEFEF}\textbf{86.7}              & \cellcolor[HTML]{EFEFEF}\textbf{3.2}              &  & \cellcolor[HTML]{EFEFEF}\textbf{99.6}               & \cellcolor[HTML]{EFEFEF}\textbf{1.5}               &  & \cellcolor[HTML]{EFEFEF}\textbf{91.4}                        & \cellcolor[HTML]{EFEFEF}\textbf{2.4}                        \\ \cline{1-2} \cline{4-4} \cline{6-7} \cline{9-10} \cline{12-13} 
                          & SELDNet \cite{Adavanne2018_JSTSP}                                &  & 0.8                                                                             &  & 51.9                                      & 9.3                                      &  & 92.8                                       & 2.1                                       &  & 73.5                                                & 4.4                                                \\
                          & SE-Resnet \cite{Han_KU_task3_report}                              &  & 10.2                                                                            &  & 80.8                                      & 3.3                                      &  & 97.9                                       & \textbf{1.4}                                      &  & 88.8                                                & 2.5                                                \\
                          & SALSA-Lite \cite{Nguyen2021SALSALiteAF}                            &  & 14.0                                                                            &  & 86.8                                      & 2.6                                      &  & 97.0                                       & 1.5                                       &  & 91.9                                                & 2.1                                                \\
\multirow{-4}{*}{Offline} & \cellcolor[HTML]{EFEFEF}FN-SSL (prop.) &  & \cellcolor[HTML]{EFEFEF}2.1                                                     &  & \cellcolor[HTML]{EFEFEF}\textbf{94.1}              & \cellcolor[HTML]{EFEFEF}\textbf{2.1}              &  & \cellcolor[HTML]{EFEFEF}\textbf{99.9}               & \cellcolor[HTML]{EFEFEF}\textbf{1.4}               &  & \cellcolor[HTML]{EFEFEF}{\color[HTML]{333333}\textbf{ 96.4}} & \cellcolor[HTML]{EFEFEF}{\color[HTML]{333333} \textbf{1.8}} \\ \hline
\label{table:simu}
\end{tabular}
\vspace{-3em}
\end{table}

\subsection{Experimental results}
\textbf{\textit{Ablation studies }} of for online SSL are presented in Table \ref{ab_study}. Block$\times N$ means $N$ full-narrow blocks are cascaded, all with the DP-IPD learning target. As the number of blocks increases, the localization performance improves, especially when the number of blocks increases from 1 to 2. This demonstrates that the full-narrow block is an effective building block for DP-IPD estimation, and deeper information can be explored by stacking the blocks. Block$\times 3$ w/o skip. shows that the performance is reduced when the skip connections are removed, and the skip connections are indeed helpful for information transmission. Additionally, we found that the skip connections can accelerate the convergence of training. As presented in Section \ref{sec:otarget}, the learning target can be directly set as location classes or coordinate to perform end-to-end location classification or regression, and Block$\times 2$ with location class. and location regress. show their performance, respectively. In location classification, 180 classes for every 1\degree  azimuth are used. Location classification achieves comparable performance with DP-IPD. Location regression performs slightly worse than DP-IPD and location classification. We found that, during training, the convergence speed of location regression is slower than the one of other training targets.
In the following comparison experiments, we use the Block$\times$3 version for the proposed network.

\textbf{\textit{Results on simulated data}} are shown in Table \ref{table:simu}. For our method, we trained both the online and offline models to fairly compare with other methods. The results on two representative acoustic conditions are reported, i.e. [SNR = 0 dB, RT60 = 1 s] (very-noisy) and [SNR = 15 dB, RT60 = 0.2 s] (relatively-quiet). AVG represents the average performance on the whole test set with SNR $\in [-5, 15]$ dB, and RT60 $\in [0.2, 1.3]$ s.
\par In the online experiments, the proposed method and SRP-DNN achieve better performance measures than Cross3D and IcoCNN. The proposed method and SRP-DNN both take as input the microphone signals and predict the direct-path localization feature, while Cross3D and IcoCNN take as input the noisy SRP-PHAT spatial spectrum and predict the source direction. Compared to processing the SRP-PHAT spatial spectrum, directly processing the microphone signals is more effective for suppressing the interferences of noise and reverberation, as the natural properties of noise and reverberation (such as the spatial-diffuseness) presented in the original signals can be better leveraged. The proposed method outperforms SRP-DNN especially for the very-noisy condition.
This indicates that the proposed full-band and narrow-band fusion network is more effective in capturing spatial cues than the full-band CRNN network used in SRP-DNN.
\begin{figure}[t]
    \centering
    \includegraphics[width=1\linewidth]{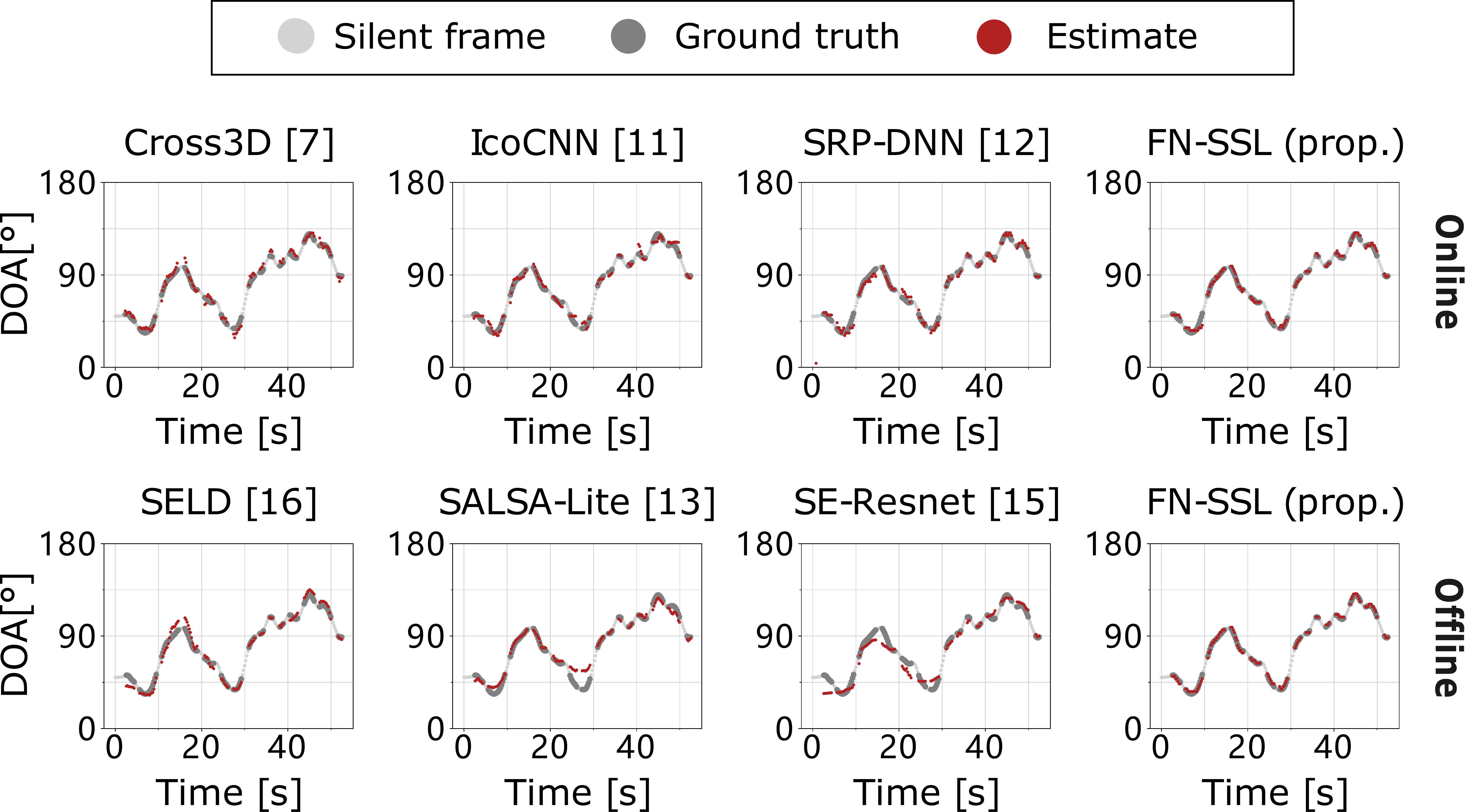}
    \vspace{-2em}
    \caption{DOA (trajectory) estimation for a LOCATA challenge recording.}
    \label{fig:locata}
    \vspace{-1em}
\end{figure}

\begin{table}[]
\caption{Results on the LOCATA dataset.}
\label{table:locata}
\vspace{-1em}
\renewcommand\arraystretch{1.1}
\tabcolsep0.025in
\footnotesize
\begin{tabular}{cccccc}
\hline
\multicolumn{3}{c}{Methods}                                           & ACC (15\degree) {[}\%{]}            & ACC (10\degree) {[}\%{]}            & MAE {[}°{]}                 \\ \hline
                          &  & Cross3D \cite{DiazGuerra2020RobustSS}                                & 97.1                         & 93.6                         & 3.9                         \\
                          &  & IcoCNN \cite{DiazGuerra2022DirectionOA}                                  & 97.5                         & 93.2                         & 3.9                         \\
                          &  & SRP-DNN \cite{Yang2022SRPDNNLD}                                & 95.8                         & 92.5                         & 4.3                         \\
\multirow{-4}{*}{Online}  &  & \cellcolor[HTML]{EFEFEF}FN-SSL (prop.) & \cellcolor[HTML]{EFEFEF}\textbf{98.2} & \cellcolor[HTML]{EFEFEF}\textbf{96.6} & \cellcolor[HTML]{EFEFEF}\textbf{2.6} \\ \cline{1-1} \cline{3-6} 
                          &  & SELDNet \cite{Adavanne2018_JSTSP}                                & 94.8                         & 90.3                         & 4.8                         \\
                          &  & SE-Resnet \cite{Han_KU_task3_report}                              & 90.6                         & 84.6                         & 6.5                         \\
                          &  & SALSA-Lite \cite{Nguyen2021SALSALiteAF}                            & 94.8                         & 91.8                         & 4.0                         \\
\multirow{-4}{*}{Offline} &  & \cellcolor[HTML]{EFEFEF}FN-SSL (prop.) & \cellcolor[HTML]{EFEFEF}\textbf{99.9} & \cellcolor[HTML]{EFEFEF}\textbf{97.7} & \cellcolor[HTML]{EFEFEF}\textbf{1.9} \\ \hline
\end{tabular}
\vspace{-2em}
\end{table}

\par In the offline experiments, the proposed method also achieves the best performance measures. SALSA-Lite, SE-Resnet, and SELDnet all process the noisy IPD (concatenated with the magnitude spectrum), with different CRNN networks. Again, suppressing the interferences of noise and reverberation from the noisy IPDs may not be as effective as suppressing them from the original microphone signals. 

\par \textbf{\textit{Results on real-world data}} are shown in Table \ref{table:locata}, and a localization example of different methods are given in Fig. \ref{fig:locata}. The models trained with the simulated dataset are directly used for testing on the LOCATA dataset. For this dataset, we have observed a consistent DOA estimation bias of about 4$\degree$ for all methods, which is possibly attributed to the annotation bias, and we compensated this bias for all methods by subtracting 4$\degree$ from their DOA estimations. 
The acoustic condition of this dataset, i.e. almost noise-free and RT60 is 0.55 s, is relatively good, and thus all methods achieve reasonable SSL results as shown in Fig. \ref{fig:locata}.  
It is worth noting that there exist some sudden turnings (zig-zags as shown in Fig. \ref{fig:locata}) in this dataset, and the offline comparison methods exhibit an over-smoothing phenomenon for these sudden turning regions. As a result, they perform even worse than the online methods.
The proposed method noticeably outperforms all the comparison methods.

\section{Conclusions}
This paper proposes the FN-SSL network, which exploits the inter-band dependencies and the inter-channel information with the dedicated full-band and narrow-band layers, respectively. FN-SSL can be performed in both online and offline manners, and used for both localization feature extraction  and end-to-end location classification or regression. Experimental results show the superiorities of the proposed FN-SSL network over other methods. In addition, the FN-SSL network trained with simulated data can be well generalized to real-world data.
\clearpage
\bibliographystyle{IEEEtran}
\bibliography{mybib}

\end{document}